\documentclass{svproc}
 
\usepackage{cite,graphicx,multicol,footmisc}
\usepackage{slashed,amsmath,amssymb}
\usepackage{cite}

\begin{document}
\mainmatter 
  
\title{Critical exponents for the valence-bond-solid transition in lattice quantum electrodynamics}
\titlerunning{Critical properties of the $O(2)$ QED$_{3}$-Gross-Neveu model}

\author{Rufus Boyack \and Joseph Maciejko} 
\institute{Department of Physics and Theoretical Physics Institute, University of Alberta, Edmonton, Alberta T6G 2E1, Canada}

\maketitle
  
\begin{abstract}
Recent sign-problem-free quantum Monte Carlo simulations of (2+1)-dimensional lattice quantum electrodynamics (QED$_3$) with $N_{f}$ flavors of fermions on the square lattice have found evidence of continuous quantum phase transitions between a critical phase and a gapped valence-bond-solid (VBS) phase for flavor numbers $N_f=4$, $6$, and $8$. We derive the critical theory for these transitions, the chiral $O(2)$ QED$_3$-Gross-Neveu model, and show that the latter is equivalent to the gauged Nambu--Jona-Lasinio model. Using known large-$N_f$ results for the latter, we estimate the order parameter anomalous dimension and the correlation length exponent for the transitions mentioned above. We obtain large-$N_f$ results for the dimensions of fermion bilinear operators, in both the gauged and ungauged chiral $O(2)$ Gross-Neveu models, which respectively describe the long-distance power-law decay of two-particle correlation functions at the VBS transition in lattice QED$_3$ and the Kekul\'e-VBS transition for correlated fermions on the honeycomb lattice.
\end{abstract}

Quantum phase transitions that involve fractionalized degrees of freedom fall outside the traditional Landau paradigm and have been the focus of much interest in recent years. The classic example is deconfined quantum critical points between conventional phases of quantum antiferromagnets~\cite{Senthil2004,Senthil2004b}, where emergent fractionalized matter fields and gauge fields appear at the critical point but are confined in the phases themselves. A class of transitions comparatively less studied, but also beyond the Landau paradigm, are transitions between phases supporting fractionalized excitations, such as different types of spin liquids, or between a fractionalized phase and a conventional phase. In the language of lattice gauge theories of quantum antiferromagnets~\cite{Wen}, where spin degrees of freedom fractionalize into emergent fermions coupled to gauge fields, these correspond to transitions between distinct deconfined phases of a lattice gauge theory, or between a deconfined phase and a confined phase, respectively. Besides their application to frustrated magnetism and elementary particle physics, lattice gauge theories may now be experimentally realized using ultracold atoms in optical lattices~\cite{zohar2015,martinez2016}, and thus constitute an important class of interacting many-body systems whose phases and phase transitions are of intrinsic interest.

Recently, sign-problem-free quantum Monte Carlo (QMC) simulations of (2+1)-dimensional lattice quantum electrodynamics (QED$_3$) with an even number $N_f$ of flavors of fermions on the square lattice~\cite{Meng2019,Meng2019b} have found evidence for a $U(1)$ deconfined phase with power-law correlations, and for continuous quantum phase transitions from this phase to conventional confined phases. For $N_f=2$, the putative $U(1)$ phase is adiabatically connected to the algebraic spin liquid~\cite{hermele2005} and the transition is towards a N\'eel antiferromagnet. This transition is described by the chiral $O(3)$ QED$_3$-Gross-Neveu (GN) model, whose universal critical properties were studied recently using both $\epsilon$-expansion~\cite{WitczakKrempa2019,Zerf2019} and large-$N_{f}$ techniques~\cite{Zerf2019}. For $N_f=4$, $6$, and $8$, the confinement transition is found to be towards a gapped valence-bond-solid (VBS) phase. The corresponding critical theory was conjectured to be of the chiral $O(2)$ QED$_3$-GN type~\cite{Meng2019}, but its critical properties have thus far not been investigated. In this paper, we establish the precise form of the critical theory, show its equivalence to the gauged Nambu--Jona-Lasinio (NJL) model~\cite{Nambu1961,klevansky1989}, and determine various critical exponents using the large-$N_f$ expansion. In both the gauged and ungauged chiral $O(2)$ GN models, we obtain new results for the scaling dimensions of fermion bilinears that respectively describe the power-law decay of certain two-particle correlation functions at the $U(1)$-VBS transition and the semimetal-to-Kekul\'e-VBS transition for interacting fermions on the honeycomb lattice~\cite{Lang2013,zhou2016,Li2017}. Critical exponents for the $U(1)$-VBS transition at $\mathcal{O}(1/N_f^2)$ in the large-$N_f$ expansion and four-loop order in the $\epsilon$-expansion will be reported in a future publication~\cite{Zerf2020}.

\section{The $U(1)$-VBS transition}

The $U(1)$ lattice gauge theory studied in Ref.~\cite{Meng2019,Meng2019b} is a quantum rotor model with fermions on the square lattice. The Hamiltonian is 
\begin{align}\label{Hlattice}
\mathcal{H}&=\frac{1}{2}JN_{f}\sum_{\langle rr'\rangle}\frac{1}{4}L^2_{rr'}-t\sum_{\alpha=1}^{N_f}\sum_{\langle rr'\rangle}(c^{\dagger}_{r\alpha}e^{i\theta_{rr'}}c_{r'\alpha}+\text{h.c.})\nonumber\\
&\quad+\frac{1}{2}KN_{f}\sum_{\square}\cos(\boldsymbol{\Delta}\times\boldsymbol{\theta}),
\end{align}
where $c_{r\alpha}^{(\dag)}$ annihilates (creates) a fermion of flavor $\alpha=1,\ldots,N_f$ on site $r$, $\langle rr'\rangle$ denotes bonds between nearest-neighbor sites $r$ and $r'$, the angular bond variable $\theta_{rr'}\in[0,2\pi)$ and the angular momentum $L_{rr'}$ are canonical conjugates, and $\boldsymbol{\Delta}\times\boldsymbol{\theta}$ denotes the lattice curl of $\theta$ around a plaquette $\square$. The magnetic coupling $K>0$ favors a background flux of $\pi$ in each plaquette. To begin, consider the fermionic part of the Hamiltonian, in the absence of gauge fluctuations ($J=0$). A gauge for the background flux can be chosen such that the Hamiltonian is 
\begin{equation}
\mathcal{H}_{0}=\sum_{\alpha=1}^{N_f}\int_{\text{BZ}}\frac{d^{2}k}{(2\pi)^2}c^{\dagger}_{\boldsymbol{k}\alpha}h(\boldsymbol{k})c_{\boldsymbol{k}\alpha},
\end{equation}
with the two-component spinor $c_{\boldsymbol{k}\alpha}=(c_{\boldsymbol{k},\alpha,A},c_{\boldsymbol{k},\alpha,B})$, where $A$ and $B$ denote the two sublattices arising from the choice of gauge, and 
\begin{equation}
h(\boldsymbol{k})=-t\left(\begin{array}{cc}
0 & f(\boldsymbol{k}) \\
f^{*}(\boldsymbol{k}) & 0
\end{array}\right), \quad f(\boldsymbol{k})=1-e^{i(k_{x}-k_{y})}+e^{-i(k_{x}+k_{y})}+e^{-2ik_{y}}.
\end{equation}
Two Dirac nodes are found at $\pm\boldsymbol{Q}=(0,\pm\frac{\pi}{2})$. Keeping only the degrees of freedom near the Dirac nodes, the low-energy Hamiltonian becomes
\begin{equation}\label{H0}
\mathcal{H}_{0}\approx v_{F}\sum_{\alpha=1}^{N_f}\sum_{\eta=\pm}\int\frac{d^{2}p}{(2\pi)^2}\chi^{\dagger}_{\alpha\eta}(\boldsymbol{p})(\mu_{1}p_{x}+\mu_{2}p_{y})\chi_{\alpha\eta}(\boldsymbol{p}),
\end{equation}
where $v_{F}=2t$. The two-component Dirac fields are defined by 
\begin{equation}
\chi_{\alpha,+}({\boldsymbol p})=\left(\begin{array}{c}
c_{{\boldsymbol{Q}}+{\boldsymbol{p}},\alpha,A}\\
c_{{\boldsymbol{Q}}+{\boldsymbol{p}},\alpha,B}
\end{array}\right),\hspace{5mm}
\chi_{\alpha,-}({\boldsymbol p})=\left(\begin{array}{c}
c_{-{\boldsymbol{Q}}+{\boldsymbol{p}},\alpha,B}\\
-c_{-{\boldsymbol{Q}}+{\boldsymbol{p}},\alpha,A}
\end{array}\right).
\end{equation}
These can be combined into $N_f$ flavors of four-component Dirac fermions $\Psi_\alpha=(\chi_{\alpha,+},\chi_{\alpha,-})^T$. We introduce the following $4\times 4$ (reducible) representation of the Euclidean Dirac algebra in 2+1 dimensions,
\begin{align}\label{Gamma_mu}
\Gamma_{\mu}=\left(\begin{array}{cc}
\tilde{\gamma}_{\mu} & 0 \\
0 & -\tilde{\gamma}_{\mu}
\end{array}\right), \quad \mu=0,1,2,
\end{align}
where $\tilde{\gamma}_\mu$ are $2\times2$ Euclidean Dirac matrices defined in terms of Pauli matrices by $(\tilde{\gamma}_{0},\tilde{\gamma}_{1},\tilde{\gamma}_{2})=(\sigma_{3},\sigma_{2},-\sigma_{1})$. Using the Dirac conjugate $\overline{\Psi}_\alpha=\Psi_\alpha^\dag\Gamma_0$, the Lagrange density for the Hamiltonian (\ref{H0}) is $\mathcal{L}_{0}=\sum_\alpha\overline{\Psi}_{\alpha}\Gamma_{\mu}\partial_{\mu}\Psi_{\alpha}$. For small but nonzero $J>0$, a Maxwell kinetic term for gauge-field fluctuations $A_\mu$ about the $\pi$-flux background is generated and $\partial_\mu$ in $\mathcal{L}_0$ is promoted to the gauge-covariant derivative $D_\mu=\partial_\mu+iA_\mu$. This results in the QED$_3$ Lagrangian, which exhibits a conformal infrared fixed point for sufficiently large $N_f$~\cite{appelquist1988}---in accordance with the critical phase observed numerically at small $J$~\cite{Meng2019,Meng2019b}. We are at present treating the $U(1)$ gauge field as noncompact; the effects of compactness due to the original lattice formulation will be discussed in Section~\ref{sec:discussion}.

For $J$ larger than some critical value $J_c$, a VBS phase with unbroken global $SU(N_f)/\mathbb{Z}_{N_f}$ symmetry is found for $N_f=4,6,8$~\cite{Meng2019,Meng2019b}. Columnar VBS order doubles the unit cell of the square lattice and spontaneously breaks the latter's $D_4$ point-group symmetry to a $D_2$ subgroup; it is represented by a time-reversal-invariant vector order parameter $\boldsymbol{V}=(V_x,V_y)$ transforming in the two-dimensional $E$ irreducible representation of $D_4$. Using the projective symmetry group approach, one can determine how gauge-invariant operators in the low-energy QED$_3$ theory transform under the microscopic lattice symmetries~\cite{Zerf2019}. Defining the two $4\times4$ Hermitian matrices
\begin{align}\label{Gamma3}
\Gamma_3=\left(\begin{array}{cc}
0 & -i \\
i & 0
\end{array}\right),\hspace{5mm}
\Gamma_5=\Gamma_{0}\Gamma_{1}\Gamma_{2}\Gamma_{3}=\left(\begin{array}{cc}
0 & 1 \\
1 & 0
\end{array}\right),
\end{align}
which square to the identity and anticommute with each other and with the Dirac matrices (\ref{Gamma_mu}), one finds that the pair of time-reversal-invariant and flavor-symmetric Dirac bilinears $\left(\sum_\alpha i\overline{\Psi}_\alpha\Gamma_5\Psi_\alpha,\sum_\alpha i\overline{\Psi}_\alpha\Gamma_3\Psi_\alpha\right)$ transform precisely in the $E$ irreducible representation of $D_4$. Furthermore, $\sum_\alpha i\overline{\Psi}_\alpha\Gamma_5\Psi_\alpha$ is odd under $x$-reflections and lattice $x$-translations and even under $y$-translations, while $\sum_\alpha i\overline{\Psi}_\alpha\Gamma_3\Psi_\alpha$ transforms oppositely. Thus one can identify $V_x\sim \sum_\alpha i\overline{\Psi}_\alpha\Gamma_5\Psi_\alpha$ and $V_y\sim \sum_\alpha i\overline{\Psi}_\alpha\Gamma_3\Psi_\alpha$. A nonzero expectation value of $\boldsymbol{V}$ corresponds to a nonzero fermion mass, in accordance with the gapped spectrum observed in the VBS phase~\cite{Meng2019b}. Note that $\boldsymbol{V}$ is Lorentz invariant since $\Gamma_3$ and $\Gamma_5$ commute with the Euclidean transformations $\exp(-\frac{i}{2}\omega_{\mu\nu}\sigma_{\mu\nu})$ where $\sigma_{\mu\nu}=\frac{i}{4}[\Gamma_\mu,
\Gamma_\nu]$.

The occurrence of a VBS phase for $J>J_c$ can be understood as arising from a short-ranged four-fermion interaction term $\sim (g^2/N_f)\boldsymbol{V}^2$ generated by gauge fluctuations at the lattice scale. Such interactions are perturbatively irrelevant at the conformal QED$_3$ fixed point, but if sufficiently strong can give rise to dynamical fermion mass generation via a quantum critical point. Decoupling this interaction term with a pair $\boldsymbol{\phi}=(\phi_1,\phi_2)$ of scalar fields and tuning to the quantum critical point, we obtain the chiral $O(2)$ QED$_3$-GN model,
\begin{align}\label{eq:Lagran}
\mathcal{L}_{\textrm{O(2) QED}_{3}\textrm{-GN}}  =  \sum_{\alpha=1}^{N_{f}}\left[\overline{\Psi}_{\alpha}\Gamma_{\mu}\left(\partial_{\mu}+\frac{e}{\sqrt{N_{f}}}iA_{\mu}\right)\Psi_{\alpha}+\frac{g}{\sqrt{N_{f}}}i\boldsymbol{\phi}\cdot\overline{\Psi}_{\alpha}\boldsymbol{M}\Psi_{\alpha}\right] +\ldots,
\end{align}
where $\boldsymbol{M}=(\Gamma_3,\Gamma_5)$, and $\ldots$ includes Maxwell and gauge-fixing terms for the gauge field, and symmetry-allowed kinetic and self-interaction terms for the scalar field $\boldsymbol{\phi}$. At the free-field fixed point, the gauge coupling $e^2$ and the Yukawa coupling $g^2$ have units of mass and are thus relevant. However, the fields have been rescaled to make explicit the fact that $e$ and $g$ appear with a suppressing factor of $1/\sqrt{N_f}$. In the large-$N_f$ limit, the physics at momenta $|q|\ll e^2,g^2$ is dominated by the coupling between fermions and soft bosonic fluctuations, i.e., the terms in square brackets in Eq.~(\ref{eq:Lagran}), and can be computed systematically in powers of $1/N_f$. Conceptually similar applications of the $1/N_f$ expansion to the chiral Ising and $O(3)$ QED$_3$-GN models can be found in Refs.~\cite{Gracey1992,Gracey1993a,GraceyAnnPhys,Alanne2018,Gracey2018,Boyack2019} and \cite{Zerf2019}, respectively. To leading (zeroth) order in this expansion, the large-$N_{f}$ scalar-field and gauge-field propagators in the infrared limit are, respectively:
\begin{align}\label{LargeNfPropag}
D_{ab}(q)=\frac{4}{g^2|q|}\delta_{ab}, \hspace{5mm} \Pi_{\mu\nu}(q)=\frac{8}{e^2|q|}\left(\delta_{\mu\nu}-\frac{q_{\mu}q_{\nu}}{q^2}\right),
\end{align}
where $a,b=1,2$. The gauge-field propagator is given in the Landau gauge.

The extra terms $\ldots$ in Eq.~(\ref{eq:Lagran}) contain a coupling of the form $\propto(\phi_1+i\phi_2)^4+\mathrm{c.c.}$, which transforms trivially under $C_4$ rotations and is thus allowed by the microscopic symmetries. Such a term is relevant at the free-field fixed point. However, Eq.~(\ref{LargeNfPropag}) implies that the scaling dimension of $\boldsymbol{\phi}$ at the chiral $O(2)$ QED$_3$-GN critical point is $\Delta_\phi=1+\mathcal{O}(1/N_f)$, thus this term is irrelevant at the $U(1)$-VBS critical point in the large-$N_f$ limit. (Other $D_4$-allowed terms are already irrelevant at the free-field fixed point.) Thus the Lagrangian (\ref{eq:Lagran}) acquires an emergent $SO(2)$ symmetry under $\Psi_\alpha\rightarrow e^{-iW\theta/2}\Psi_\alpha$, $\phi_a\rightarrow R_{ab}(\theta)\phi_b$, where $W=-i\Gamma_3\Gamma_5$ and $R(\theta)$ is the $SO(2)$ matrix for a rotation through angle $\theta$.

\section{The gauged NJL model and critical exponents}

The NJL model was originally introduced as a toy model of chiral symmetry breaking and dynamical mass generation in high-energy physics~\cite{Nambu1961}. Its gauged version~\cite{klevansky1989} is described by the Lagrangian
\begin{align}
\label{eq:NJL_Lagran}
\mathcal{L}_{\mathrm{NJL}}  =  \sum_{\alpha=1}^{N_{f}}\left[\overline{\psi}_{\alpha}\gamma_{\mu}\left(\partial_{\mu}+\frac{e}{\sqrt{N_{f}}}iA_{\mu}\right)\psi_{\alpha}+\frac{g}{\sqrt{N_{f}}}\overline{\psi}_{\alpha}\left(\phi_{1}+i\phi_{2}\gamma_{5}\right)\psi_{\alpha}\right]+\ldots,
\end{align}
where $\psi_\alpha$ are four-component Dirac spinors, and, as previously, $\ldots$ denotes terms not involving fermions which, besides a gauge-fixing term, are irrelevant in the large-$N_f$ limit of interest to us. We now show that the gauged NJL model is entirely equivalent to the chiral $O(2)$ QED$_3$-GN model (\ref{eq:Lagran}). 
Define the gamma matrices in Eq.~(\ref{eq:NJL_Lagran}) in terms of those in Eq.~(\ref{Gamma_mu}-\ref{Gamma3}) by $\gamma_{\mu}=i\Gamma_{\mu}\Gamma_{3}$, $\mu=0,1,2$ and $\gamma_{5}=-i\Gamma_{3}\Gamma_{5}$, and in addition define $\Psi_\alpha=\psi_\alpha$ and $\overline{\psi}_\alpha=\Psi^{\dagger}_\alpha\gamma_{0}$. The Hermitian matrices $\gamma_\mu$ and $\gamma_5$ obey the usual Euclidean Dirac algebra (i.e., they anticommute with each other and square to the identity). Using these gamma matrices, the gauged NJL Lagrangian (\ref{eq:NJL_Lagran}) becomes equal to the chiral $O(2)$ QED$_3$-GN Lagrangian (\ref{eq:Lagran}). The emergent $SO(2)$ symmetry of the latter is identified with the invariance of the former under $U(1)$ chiral transformations $\psi_\alpha\rightarrow e^{-i\gamma_5\theta/2}\psi_\alpha$, with a concomitant rotation of the scalar field $\boldsymbol{\phi}=(\phi_1,\phi_2)$.

If the gauge field is absent, this also establishes the equivalence between the ungauged NJL model and the chiral $O(2)$ GN model. The latter describes the semimetal-to-Kekul\'e-VBS transition for interacting fermions on the honeycomb lattice~\cite{Lang2013,zhou2016,Li2017}. The $D_6$ point-group symmetry of the honeycomb lattice allows for a term of the form $\propto(\phi_1+i\phi_2)^3+\mathrm{c.c.}$ in the critical Lagrangian, which is marginal in the $N_f=\infty$ limit at the chiral $O(2)$ GN fixed point. However, a renormalization-group analysis in the large-$N_f$ limit shows that the $\mathcal{O}(1/N_f)$ correction renders this term irrelevant~\cite{Li2017}. QMC simulations of the joint probability distribution $P(\phi_1,\phi_2)$ of the two components of the VBS order parameter also support the emergent $SO(2)$ symmetry at the critical point~\cite{Li2017}.

The critical points of the gauged and ungauged NJL models are strongly coupled (2+1)-dimensional conformal field theories characterized by a spectrum of scaling dimensions that correspond to universal critical exponents. Some of these exponents have already been computed in the $1/N_{f}$ expansion in general $d$ spacetime dimensions~\cite{Gracey1993,Gracey1993b,Gracey1994,Gracey1994b}. The order-parameter anomalous dimension for $d=3$ is
\begin{align}
\text{chiral $O(2)$ QED$_3$-GN}: \quad \eta_{\phi}&=1+\frac{56}{3\pi^2N_{f}}+\mathcal{O}(1/N_f^2), \\
\text{chiral $O(2)$ GN}: \quad \eta_{\phi}&=1-\frac{8}{3\pi^2N_{f}}+\frac{544}{27\pi^{4}N_{f}^2}+\mathcal{O}(1/N_f^3),
\end{align}
and is related to the scalar-field scaling dimension by $\Delta_{\phi}=\frac{1}{2}\left(1+\eta_{\phi}\right)$. The inverse correlation-length exponent is
\begin{align}
\text{chiral $O(2)$ QED$_3$-GN}: \quad \nu^{-1}&=1-\frac{80}{3\pi^2N_{f}}+\mathcal{O}(1/N_f^2), \\
\text{chiral $O(2)$ GN}: \quad \nu^{-1}&=1-\frac{16}{3\pi^2N_{f}}+\frac{8(364+27\pi^{2})}{27\pi^{4}N_{f}^2}+\mathcal{O}(1/N_f^3),
\end{align}
and is related to the scaling dimension of the $\boldsymbol{\phi}^2$ operator by $\Delta_{\phi^2}=3-\nu^{-1}$.

\section{Fermion bilinear scaling dimensions}

The scaling dimension $\Delta_\phi$ characterizes the power-law decay at long distances of the microscopic VBS correlation function $\langle\mathcal{O}_\text{VBS}(r)\mathcal{O}_\text{VBS}(r')\rangle\sim|r-r'|^{-2\Delta_\phi}$ at the $U(1)$-VBS critical point of the lattice gauge theory (\ref{Hlattice}), where $\mathcal{O}_\text{VBS}(r)$ can be chosen as $(-1)^x\sum_A{S}_A(r)S_A(r+\hat{x})$ or $(-1)^y\sum_A{S}_A(r)S_A(r+\hat{y})$, describing columnar VBS order in the $x$ and $y$ directions, respectively. Here $S_A(r)=\sum_{\alpha\beta}c_{r\alpha}^\dag T_A^{\alpha\beta}c_{r\beta}$ denotes the $SU(N_f)$ spin operator, where $T_A$, $A=1,\ldots,N_f^2-1$ is a Hermitian generator of the $SU(N_f)$ Lie algebra. The corresponding thermodynamic susceptibility $\chi_\text{VBS}(\boldsymbol{q})\sim|\boldsymbol{q}|^{2\Delta_\phi-3}\sim|\boldsymbol{q}|^{-(2-\eta_\phi)}$, which can be computed in QMC, diverges as $\boldsymbol{q}\rightarrow 0$, signaling the onset of VBS order. In the $U(1)$ phase itself, which is a critical phase, other gauge-invariant observables such as the staggered density operator $\mathcal{O}_\textrm{CDW}(r)=(-1)^{x+y}\sum_\alpha c_{r\alpha}^\dag c_{r\alpha}$, the staggered $SU(N_f)$ spin $\mathcal{O}_A(r)=(-1)^{x+y}S_A(r)$, and a quantum anomalous Hall mass operator $\mathcal{O}_\text{QAH}(r)$ defined in Ref.~\cite{Zerf2019} also exhibit universal power-law correlations characterized by (non-diverging) susceptibilities $\chi_\mathcal{O}(\boldsymbol{q})\sim|\boldsymbol{q}|^{2\Delta_\mathcal{O}-3}$~\cite{hermele2005}, which are also in principle accessible in QMC. Such susceptibilities remain power law at the $U(1)$-VBS critical point, but with different exponents characterizing the conformal field theory associated with the chiral $O(2)$ QED$_3$-GN fixed point as opposed to that of the pure-QED$_3$ fixed point. Detecting a change in these exponents numerically upon approach to the critical point $J\rightarrow J_c^-$ would be a signature of the new universality class discussed in the present paper.

At the $U(1)$-VBS critical point, the microscopic observables above correspond in the long-wavelength limit to Lorentz-invariant fermion bilinears in the chiral $O(2)$ QED$_3$-GN field theory (\ref{eq:Lagran}): the flavor-singlet, time-reversal-even bilinear $\overline{\Psi}\Psi\sim\mathcal{O}_\textrm{CDW}$, the flavor-adjoint, time-reversal-even bilinear $\overline{\Psi}T_A\Psi\sim\mathcal{O}_A$, and the flavor-singlet, time-reversal-odd bilinear $i\overline{\Psi}\Gamma_3\Gamma_5\Psi\sim\mathcal{O}_\text{QAH}$. We have computed the scaling dimensions of these bilinears at $\mathcal{O}(1/N_f)$ in the large-$N_f$ expansion by adapting the methods used in Ref.~\cite{Boyack2019} for the chiral Ising QED$_3$-GN model, accounting for the matrix structure in the Yukawa vertex and the anticommutation properties of $\Gamma_{3}$ and $\Gamma_{5}$. We obtain:
\begin{align}
\text{chiral $O(2)$ QED$_3$-GN}: \quad \Delta{}_{\overline{\Psi}\Psi}=\Delta{}_{\overline{\Psi}T_{A}\Psi}&=2-\frac{40}{3\pi^{2}N_{f}}+\mathcal{O}(1/N_f^2), \\
\Delta{}_{i\overline{\Psi}\Gamma_{3}\Gamma_{5}\Psi}&=2+\frac{80}{3\pi^{2}N_{f}}+\mathcal{O}(1/N_f^2),\\
\text{chiral $O(2)$ GN}: \quad \Delta{}_{\overline{\Psi}\Psi}=\Delta{}_{\overline{\Psi}T_{A}\Psi}&=2-\frac{8}{3\pi^{2}N_{f}}+\mathcal{O}(1/N_f^2),\\
\Delta{}_{i\overline{\Psi}\Gamma_{3}\Gamma_{5}\Psi}&=2+\frac{16}{3\pi^{2}N_{f}}+\mathcal{O}(1/N_f^2).
\end{align}

\section{Discussion}
\label{sec:discussion}

In Tables \ref{tab:O2QEDGN} and \ref{tab:O2GN}, we evaluate the previous expressions at values of $N_f$ currently accessible to QMC simulations to obtain estimates of critical exponents at the $U(1)$-VBS and semimetal-to-Kekul\'e-VBS transitions, respectively. In Table \ref{tab:O2GN} we also provide the values of $\eta_\phi$ and $\nu$ already obtained from QMC simulations~\cite{Li2017}. For $\nu$, a better agreement with QMC results is found than with a previously used renormalization-group approach~\cite{Li2017}, while the opposite is true for $\eta_\phi$.

\begin{table}[t]
\centering
\begin{tabular}{|l||c|c|c|c|c|c|}
      \hline
    &                            $\eta_\phi$  & $1/\nu$   & $\nu$     & $\Delta_{\overline{\Psi}\Psi}$    & $\Delta_{i\overline{\Psi}\Gamma_{3}\Gamma_{5}\Psi}$\\
       \hline
$N_{f}=4$                   & 1.473      & 0.3245     & 3.081   & 1.662                                          & 2.675  \\
      \hline
$N_{f}=6$                   & 1.315      & 0.5497     & 1.819  & 1.775                                           & 2.450  \\
      \hline 
$N_{f}=8$                   &  1.236   & 0.6623       & 1.510 & 1.831                                            & 2.338 \\
     \hline
\end{tabular}
      \caption{Large-$N_{f}$ critical exponents for the chiral $O(2)$ QED$_3$-GN model.}
      \label{tab:O2QEDGN}
\end{table}

\begin{table}[t]
\centering
\begin{tabular}{|l||c|c|c|c|c|c|c|c|}
      \hline
    &                           $\eta_\phi$ & $\eta_\phi$ (QMC)  & $1/\nu$   & $\nu$ & $\nu$ (QMC)     & $\Delta_{\overline{\Psi}\Psi}$    & $\Delta_{i\overline{\Psi}\Gamma_{3}\Gamma_{5}\Psi}$\\
       \hline
$N_{f}=2$                   & 0.9166  & 0.71(3)    & 1.209    & 0.8270   & 1.06(5)     & 1.865                                            & 2.270  \\
      \hline
$N_{f}=3$                   & 0.9329 & 0.78(2)    & 1.033    & 0.9681   & 1.07(4)     & 1.910                                          & 2.180  \\
      \hline
$N_{f}=4$                   & 0.9454  & 0.80(4)    & 0.9848  & 1.015   & 1.11(3)       & 1.932                                          & 2.135  \\
      \hline
$N_{f}=5$                   & 0.9542  & 0.85(4)    & 0.9686  & 1.032   & 1.07(2)       & 1.946                                          & 2.108  \\
      \hline 
$N_{f}=6$                   & 0.9607 & 0.87(4)  & 0.9632    & 1.038    & 1.06(3)       & 1.955                                        & 2.090 \\
     \hline
\end{tabular}
      \caption{Large-$N_{f}$ and QMC critical exponents for the chiral $O(2)$ GN model.}
      \label{tab:O2GN}
\end{table}

Our discussion of the $U(1)$-VBS transition has thus far ignored the compactness of the $U(1)$ gauge field, which may cause monopole (instanton) proliferation. In the large-$N_f$ limit, the scaling dimension of the smallest symmetry-allowed monopole operator at the conformal QED$_3$ fixed point is $\Delta_M=0.53N_f-0.0383+\mathcal{O}(1/N_f)$~\cite{Pufu2014}. This suggests that for $N_f=6$ and $8$, $\Delta_M>3$ and monopoles are irrelevant, while for $N_f=4$ the smallest monopole is relevant; however, at such values of $N_f$, subleading corrections in the $1/N_f$ expansion may be significant. As with the chiral $O(3)$ QED$_3$-GN model~\cite{WitczakKrempa2019}, at the chiral $O(2)$ QED$_3$-GN fixed point the scaling dimension of the smallest monopole operator is expected to grow linearly with $N_f$ at leading order in $1/N_f$ but with a different coefficient than at the conformal QED$_3$ fixed point. Should this coefficient be sufficiently small, the $U(1)$-VBS critical point may be destabilized at sufficiently small $N_f$, resulting in a first-order transition. The QMC results~\cite{Meng2019,Meng2019b} however suggest that a continuous $U(1)$-VBS transition persists with increasing $N_f\geq 4$ but is simply pushed to larger values of $J_c$. While the critical value of $N_f$ above which monopole operators are irrelevant is not precisely known, the numerical observation of a continuous transition suggests that either monopoles are in fact irrelevant for $N_f\geq 4$, or the crossover length scale $L_*\sim a g_0^{-1/(3-\Delta_M)}$ beyond which they proliferate, where $a$ is the lattice constant and $g_0$ the bare monopole fugacity, is much larger than the system sizes currently accessible in QMC.

{\it Acknowledgements.} We thank the CRM and the QTS-XI committee for organizing this excellent conference. We thank J. A. Gracey, P. Marquard, and N. Zerf for collaboration on related topics, \'E. Dupuis, S. Giombi, I. F. Herbut, I. R. Klebanov, Z. Y. Meng, A. Penin, M. M. Scherer, and W. Witczak-Krempa for useful discussions, and NSERC, CIFAR, the University of Alberta's Theoretical Physics Institute (TPI), and the CRC program for financial support.

\end{document}